\documentclass[a4,article,prl,nofootinbib, twocolumn]{revtex4-1}

\usepackage{graphicx}
\usepackage{dcolumn}
\usepackage{bm}
\usepackage{color}
\usepackage{amsmath}%
\usepackage{amsfonts}%
\usepackage{amssymb}%
\usepackage{breqn}
\usepackage{fixfoot}

\makeatletter
\let\cat@comma@active\@empty
\makeatother

\begin{document}
\title{Self-Organization and Heating by Inward Diffusion in Magnetospheric Plasmas}
\author{N. Sato, Z. Yoshida, and Y. Kawazura}
\affiliation{Graduate School of Frontier Sciences, The University of Tokyo,
Kashiwa, Chiba 277-8561, Japan}
\date{\today}

\begin{abstract}
\noindent Through the process of inward diffusion, a strongly localized clump of plasma is created in a magnetosphere.
The creation of the density gradient, instead of the usual flattening by a diffusion process, can be explained by the
topological constraints given by the adiabatic invariants of magnetized particles \cite{YosFol,Sato}.
After developing a canonical formalism for the standard guiding center dynamics in a dipole magnetic field,
we complete our attempt to build a statistical mechanics on a constrained phase space by discussing the construction principles of the associated diffusion operator. We then 
investigate the heating mechanism associated with inward diffusion:
as particles move toward regions of higher magnetic field, they experience preferential
heating of the perpendicular (with respect to the magnetic field) temperature in order to
preserve the magnetic moment. A relationship between conservation of bounce action and temperature
isotropy emerged. We further show that this behavior is scaled by the diffusion parameter 
of the Fokker-Planck equation. These results are confirmed by numerical simulations.
\end{abstract}


\maketitle

\begin{normalsize}

\section{Introduction}

\noindent Magnetospheres \cite{Voy1,Voy2,CRPWSE} are the paradigm of systems that seemingly exhibit deviation from the classical understanding of physical principles. Through the process of inward diffusion (or radial/up-hill diffusion), magnetized particles build-up an heterogeneous structure along the equator of the dipole magnetic field \cite{Bir,Schulz,YosVortex,Box}. The resulting formation of a radiation belt seems to contradict the second law of thermodynamics.      

Recently, self-organization of such plasma structures was put in the perspective of a topologically constrained particle dynamics arising from adiabatic invariants of magnetized particles \cite{YosRT1,YosFol}. Further investigations have shown the consistency of the theory with the observed rigidly rotating thermal equilibrium of non-neutral plasmas in a magnetospheric configuration \cite{Sato2}. Additional research has led to the formulation of a diffusion operator on an appropriate phase space reflecting the geometrical constraints imposed by adiabatic invariants: a flat distribution in these proper coordinates yields a non-uniform density in the Cartesian reference system \cite{Sato}. 

The interest in the mechanism of magnetospheric self-organization stems from the wide areas of both theoretical and applied physics involved in its understanding. A well-known mathematical issue is whether variable reduction resulting from topological constraints leads to a Lagrangian/Hamiltonian formalism \cite{Cary} on the contracted space. This problem relates to the type of dynamical constraints (holonomic or not) affecting the system \cite{Bloch,Schaft,Bates}. 
From the point of view of plasma physics, it is desirable to precisely understand the mechanism behind dipole confinement, especially for nuclear fusion and matter/antimatter confinement applications \cite{Has, Pedersen}.
Planetary magnetospheres are also the object of intense astronomical observations \cite{Per,Per2,Schippers} for their relevance in space weather forecasting. 

Recent measurements in the RT-1 machine have reported the role of inward diffusion in heating-up the confined plasma concurrently with breaking of the second adiabatic invariant \cite{Kaw}. This suggests that magnetized particles not only accumulate to form a radiation belt, but also undergo a progressive acceleration due to the topological constraint affecting the non-spatial part of phase-space. Magnetospheric self-organization works both as a magnetic trap and a heating device. 

In the present paper, extending the diffusion model developed in \cite{Sato}, we investigate this
acceleration mechanism. 

In section II we write down the guiding center equations of motion under investigation. In section III we separate cyclotron motion and reduce the dynamical space to a foliation. The equations of motion are then rewritten in terms of the novel coordinates spanning it. We further show that the reduced flow is measure preserving, even if not written in terms of canonically paired variables. In section IV canonical variables are constructed by deriving the associated symplectic $2$-form. In section V we obtain the corresponding Fokker-Planck equation (FPE).  
In sections VI and VII we make a correspondence between the parameters of the FPE for inward diffusion and the time-scale/strength of fluctuations. The relation of diffusion with preservation of bounce action is also shown. In section VIII we study the plasma temperature profile: our equation predicts a higher anisotropy for faster, stronger diffusion, with the corresponding destruction of the second adiabatic invariant. The phenomenon of preferential heating of the normal temperature is also reproduced. Section IX is for the conclusions. 

Finally, we will exploit stochastic calculus to handle the change of variables from magnetic coordinates on the foliation to the oridnary Cartesian reference system \cite{Gar,Ris}.

\section{Guiding Center Dynamics in a Dipole Magnetic Field}

\noindent We study the dynamics of a gyrating particle obeying the guiding center equations:

\begin{subequations}
\begin{align}
&\dot{v}_{\parallel}=-\frac{1}{m}\left(\mu B+e\phi\right)_{l}+v_{\parallel}\boldsymbol{v}_{\boldsymbol{E}\times\boldsymbol{B}}\cdot\boldsymbol{k}\label{ap}\\
&\boldsymbol{v}=\boldsymbol{v}_{\parallel}+\boldsymbol{v}_{\boldsymbol{E}\times\boldsymbol{B}}+\boldsymbol{v}_{\nabla B}+\boldsymbol{v}_{\boldsymbol{k}}\label{vd}
\end{align}
\end{subequations}

\noindent Here, $\mu$ is the magnetic moment, $e$ the electric charge, $\boldsymbol{B}=\nabla\psi\times\nabla\theta$ the magnetic field, $\psi$ the flux function, $\theta$ the toroidal angle, $\phi$ the electric potential, $\boldsymbol{v}_{\parallel}$ the velocity along magnetic field lines, $l$ the length along a field line, $\boldsymbol{k}=\partial_{l}\left(\boldsymbol{B}/B\right)=\partial_{l}\partial_{l}=\partial^{2}_{l}$ the curvature of the magnetic field, $\boldsymbol{v}$ the particle velocity, $\boldsymbol{v}_{\boldsymbol{E}\times\boldsymbol{B}}$, $\boldsymbol{v}_{\nabla B}$, and $\boldsymbol{v}_{\boldsymbol{k}}$ the $\boldsymbol{E}\times\boldsymbol{B}$, $\nabla B$, and curvature drifts, and the pedix notation is used for derivation, except when differently specified.

\section{Separation of Cyclotron Motion}

\noindent Conservation of magnetic moment $\mu$ implies that we can separate the canonical pair $\left(\theta_{c},\mu\right)$, with $\theta_{c}$ the phase of the gyration, from the full particle dynamics invariant measure:
 
\begin{dmath}
dxdydzdp_{x}dp_{y}dp_{z}=
m^{2}dldv_{\parallel}d\theta d\psi \times d\theta_{c}d\mu\label{IM}
\end{dmath} 

\noindent The $\times$ symbol 
underlines the operated separation. Here, $\left(\boldsymbol{x},\boldsymbol{p}\right)$ are the usual canonical variables and we used the fact that $\left\lvert\nabla l\cdot\nabla\psi\times\nabla\theta\right\rvert=B$, $p_{\parallel}=mv_{\parallel}$, $p_{\perp}=mv_{\perp}=m(v_{c}cos\theta_{c}+\boldsymbol{v}_{d}\cdot\partial_{\perp})$, and $p_{\theta}=m(v_{c}sin\theta_{c}+\boldsymbol{v}_{d}\cdot \partial_{\theta}/\lvert \partial_{\theta}\rvert)+e\psi/r$, with $2\mu B=mv_{c}^{2}$ and $\boldsymbol{v}_{d}=\boldsymbol{v}-\boldsymbol{v}_{c}$.
In our notation $\partial_{l}=\boldsymbol{B}/B$ is the unit tangent vector to field lines, $\partial_{\perp}=\nabla\psi/\lvert\nabla\psi\rvert$ the unit normal vector to field lines, $\partial_{\theta}/\lvert\partial_{\theta}\rvert$ the unit vector in the toroidal direction. Thanks to the separation, the actual dynamical variables are reduced to $4$: $\left(l,v_{\parallel},\theta,\psi\right)$. Let us translate equations (\ref{ap}) and (\ref{vd}) in these new coordinates:

\begin{subequations}
\begin{align}
&\dot{l}=\nabla l\cdot\boldsymbol{v}=v_{\parallel}-q\phi_{\theta}\label{ldot}\\
&\dot{v}_{\parallel}=-\frac{1}{m}\left(\mu B+e\phi\right)_{l}+v_{\parallel}q_{l}\phi_{\theta}\label{vdot}\\
&\dot{\theta}=\nabla\theta\cdot\boldsymbol{v}=\left(\partial_{\psi}+q\partial_{l}\right)\left(\frac{\mu}{e} B+\phi\right)
-\frac{m}{e}v^{2}_{\parallel}q_{l}\label{thetadot}\\
&\dot{\psi}=\nabla\psi\cdot\boldsymbol{v}=-\phi_{\theta}\label{psidot}
\end{align}
\end{subequations}

\noindent where we introduced the quantity $q=-\partial_{l}\cdot\partial_{\psi}=\nabla l\cdot\nabla\psi/(\lvert\nabla\psi\rvert)^{2}$. One can verify that the above equations are already measure preserving: if $X=\left(\dot{l},\dot{v_{\parallel}},\dot{\theta},\dot{\psi}\right)$ is the dynamical flow and $vol^{4}=dl\wedge dv_{\parallel}\wedge d\theta\wedge d\psi$, we have

\begin{equation}
\mathfrak{L}_{X}vol^{4}=div(X)vol^{4}=
0\label{measure}
\end{equation}

\noindent This result is important, because the property required to introduce a consistent definition of entropy on the foliation (\textit{leaf-entropy}) is a preserved volume element on the contracted space. However, as we will see later on, $\left(l,v_{\parallel},\theta,\psi\right)$ are not canonical variables. We also underline that a lacking invariant measure is not an impediment when building a statistical mechanics on the separated leaf.

\section{Construction of Canonical Variables}

\noindent In order to obtain canonical variables, we look for the two-form $\omega$ describing the dynamical flow $X$:

\begin{equation}
i_{X}\omega=-dH\label{omega2}
\end{equation}

\noindent where $i$ is a contraction and $H$ the energy of the system.
To calculate $\omega$, we need $H$. We expect the energy function:

\begin{equation}
H=\frac{m}{2}v_{\parallel}^{2}+\mu B+e\phi
\end{equation} 

\noindent to be constant under the flow generated by $X$. This is because guiding center drifts
result from neglecting the particle mass in the equations of motion perpendicular to the magnetic field.
This means that drifts do not contribute with their kinetic energy to $H$. Indeed,
one can verify that $\mathcal{L}_{X}H=i_{X}dH=\dot{H}=0$.
Solving equation (\ref{omega2}) for $\omega$, one obtains:

\begin{equation}
\omega=dv_{\parallel}\wedge dl+v_{\parallel}q_{l}d\psi\wedge dl+d\psi\wedge d\theta+qd\psi\wedge dv_{\parallel}\label{omega2b}
\end{equation}

\noindent For the system to be Hamiltonian, $\omega$ must be non-degenerate and exact. That the determinant of $\omega$ is different from zero is straightforward, thus this two-form is non-degenerate. Let us see how $\omega$ is also exact, i.e. $\omega=d\lambda$, for some one-form $\lambda$. From (\ref{omega2b}):

\begin{equation}
\omega=
dv_{\parallel}\wedge dl+d\psi\wedge d\left(\theta+qv_{\parallel}\right)=d\left(v_{\parallel}dl+\psi d\eta\right)\label{omega2c}
\end{equation}

\noindent where we introduced the new variable $\eta=\theta+qv_{\parallel}$. 
(\ref{omega2c}) shows that $\omega$ is exact, with $\lambda=v_{\parallel}dl+\psi d\eta$. It follows that $\left(l,v_{\parallel}\right)$ and $\left(\psi,\eta\right)$ form two canonical pairs. 

\section{Fokker-Planck Equation on a Foliated Phase Space}

\noindent To investigate the heating mechanism of inward diffusion, we will study the time evolution of the
guiding centers' distribution function by solving an appropriate FPE.
The procedure required to obtain it is given in \cite{Sato}.
Here, we extend this equation by adding the toroidal angle $\theta$ and by introducing curvature effects.

In order to derive the desired transport equation we need to convert equations (\ref{ldot}), (\ref{vdot}), (\ref{thetadot}), and (\ref{psidot}) to a system of stochastic differential equations. 
To understand the conversion procedure, it is useful to give a formal discussion of the exploited working hypothesis
(${}_{e}\langle\,\,\rangle$ stands for ensemble average):

\noindent $1$. We assume overall charge neutrality ${}_{e}\langle\boldsymbol{E}\rangle=0$.

\noindent $2$. The drifting velocity $\boldsymbol{v}_{\boldsymbol{E}\times\boldsymbol{B}}$,
belonging to the tangent bundle $TM$, is a random processes with null ensemble average:

\begin{equation}
{}_{e}\left\langle \boldsymbol{v}_{\boldsymbol{E}\times\boldsymbol{B}}\right\rangle=\boldsymbol{0}\label{ens}
\end{equation} 

\noindent This requirement, which physically states that there are no deterministic currents associated with the fluctuations, is important when applying the change of variables formula
of stochastic calculus. A formal justification of equation (\ref{ens}) follows by noting that charge neutrality implies: 
%
$\boldsymbol{0}={}_{e}\langle\boldsymbol{v}_{\boldsymbol{E}\times\boldsymbol{B}}\rangle\times\boldsymbol{B}+{}_{e}\langle\boldsymbol{E}\rangle={}_{e}\langle\boldsymbol{v}_{\boldsymbol{E}\times\boldsymbol{B}}\rangle\times\boldsymbol{B}$.

\noindent $3$. Electromagnetic fluctuations are such that the ergodic hypothesis is satisfied on the leaf $\left(l,v_{\parallel},\theta,\psi\right)$ obtained by separating cyclotron motion $\left(\theta_{c},\mu\right)$ from the original invariant measure (\ref{measure}). 
Physically, this requirement is just saying that the first adiabatic invariant is such robust that the actual phase-space accessible to the magnetized rings is the symplectic sub-manifold $dldv_{\parallel}d\theta d\psi$. Consequently, we build a (possibly constrained) statistical mechanics on this reduced invariant measure by introducing Wiener processes.
In mathematical terms we are asking the `leaf' electric field to satisfy: 

\begin{equation}
\mathcal{E}=
-d\phi=
\frac{m}{e}D_{\parallel}^{1/2}\Gamma_{\parallel}dl+D_{\perp}^{1/2}\Gamma_{\perp}d\theta+D_{\theta}^{1/2}\Gamma_{\theta}d\psi\label{Eleaf} 
\end{equation}

\noindent Here $mD_{\parallel}^{1/2}\Gamma_{\parallel}/e=-\phi_{l}$, $D_{\perp}^{1/2}\Gamma_{\perp}=-\phi_{\theta}$, and $D_{\theta}^{1/2}\Gamma_{\theta}=-\phi_{\psi}$ are Gaussian white noises with $\Gamma dt=dW$. 
The parameters $D_{\parallel}$, $D_{\perp}$, and $D_{\theta}$ are constants  scaling the strength of diffusion.
Let us calculate the expression of the electric field $\boldsymbol{E}$ in the usual Cartesian coordinates:

\begin{dmath}
\boldsymbol{E}=
\frac{m}{e}D_{\parallel}^{1/2}\Gamma_{\parallel}\nabla l+D_{\perp}^{1/2}\Gamma_{\perp}\nabla\theta+D_{\theta}^{1/2}\Gamma_{\theta}\nabla\psi\label{E}
\end{dmath}

\noindent Using equation (\ref{E}), the $\boldsymbol{E}\times\boldsymbol{B}$ velocity becomes:

\begin{dmath}
\boldsymbol{v}_{\boldsymbol{E}\times\boldsymbol{B}}=
D_{\perp}^{1/2}\frac{\Gamma_{\perp}}{rB}\partial_{\perp}-\left(D_{\theta}^{1/2}\Gamma_{\theta}+\frac{m}{e}qD_{\parallel}^{1/2}\Gamma_{\parallel}\right)\partial_{\theta}\label{vexb}
\end{dmath}

\noindent Note that this expression is consistent with equation (\ref{ens}).
It is straightforward to deduce the spatial displacements $dX_{\theta}$ and $dX_{\perp}$ along $\partial_{\theta}/\lvert\partial_{\theta}\rvert$ and $\partial_{\perp}$ caused by (\ref{vexb}) (upper-case letters are used to specify random variables): 

\begin{subequations}
\begin{align}
&dX_{\theta}=
-r\left(D_{\theta}^{1/2}dW_{\theta}+\frac{m}{e}qD_{\parallel}^{1/2}dW_{\parallel}\right)\\
&dX_{\perp}=
\frac{D_{\perp}^{1/2}}{rB}dW_{\perp}
\end{align}
\end{subequations}

\noindent In order to obtain the desired stochastic differential equations, it is sufficient to apply the change of variables formula of stochastic calculus, by noting that the changes $d\Psi$ and $dL$ are affected by $dX_{\perp}$,
the change $d\Theta$ by $dX_{\theta}$, and the change $dV_{\parallel}$ by $dW_{\parallel}$ and $dW_{\perp}$:  

\begin{widetext}
\begin{subequations}
\begin{align}
&dL=\left\{v_{\parallel}+\mathfrak{C}_{l}+\left(\frac{1}{2}-\alpha\right)D_{\perp}\left[\left(\partial_{\psi}+q\partial_{l}\right)q+q\left(\partial_{\psi}+q\partial_{l}\right)\ln(rB)\right]\right\}dt+qD_{\perp}^{1/2}dW_{\perp}\label{dL}\\
&dV_{\parallel}=-\left(\frac{\mu}{m}B_{l}+\gamma v_{\parallel}-\mathfrak{C}_{v_{\parallel}}
\right)dt+D_{\parallel}^{1/2}dW_{\parallel}-D^{1/2}_{\perp}v_{\parallel}q_{l}dW_{\perp}\label{dV}\\
&d\Theta=\left[\frac{\mu}{e}\left(\partial_{\psi}+q\partial_{l}\right)B+\mathfrak{C}_{\theta}-\frac{m}{e}v^{2}_{\parallel}q_{l}\right]dt-D_{\theta}^{1/2}dW_{\theta}-\frac{m}{e}qD_{\parallel}^{1/2}dW_{\parallel}\label{dTheta}\\
&d\Psi=\left[D_{\perp}\left(\frac{1}{2}-\alpha\right)\left(\partial_{\psi}+q\partial_{l}\right)\ln\left(rB\right)+\mathfrak{C}_{\psi}\right]dt+D_{\perp}^{1/2}dW_{\perp}\label{dPsi}
\end{align}
\end{subequations}
\end{widetext}





\noindent The last step is to translate these equations 
into a FPE. The result is:

\begin{widetext}
\begin{dmath}
\frac{\partial P}{\partial t}=-\frac{\partial}{\partial l}\left\{v_{\parallel}+\left(\frac{1}{2}-\alpha\right)D_{\perp}\left[\left(\partial_{\psi}+q\partial_{l}\right)q+q\left(\partial_{\psi}+q\partial_{l}\right)\ln(rB)\right]+\mathfrak{C}_{l}\right\}P
+\frac{\partial}{\partial v_{\parallel}}\left(\frac{\mu}{m}B_{l}+\gamma v_{\parallel}\right-\mathfrak{C}_{v_{\parallel}})P-\frac{\partial}{\partial\theta}\left[\frac{\mu}{e}\left(\partial_{\psi}+q\partial_{l}\right)B+\mathfrak{C}_{\theta}-\frac{m}{e}v^{2}_{\parallel}q_{l}\right]P
-\frac{\partial}{\partial\psi}\left[D_{\perp}\left(\frac{1}{2}-\alpha\right)\left(\partial_{\psi}+q\partial_{l}\right)\ln\left(rB\right)+\mathfrak{C}_{\psi}\right]P
+\frac{1}{2}D_{\perp}\frac{\partial^{2}}{\partial l^{2}}q^{2}P+\frac{1}{2}D_{\parallel}\frac{\partial^{2} P}{\partial v_{\parallel}^{2}}+\frac{1}{2}D_{\theta}\frac{\partial^{2} P}{\partial\theta^{2}}
-\frac{m}{e}D_{\parallel}\frac{\partial^{2}}{\partial v_{\parallel}\partial\theta}qP+
\frac{m^{2}}{2e^{2}}D_{\parallel}\frac{\partial^{2}}{\partial\theta^{2}}q^{2}P+
\frac{1}{2}D_{\perp}\frac{\partial^{2} P}{\partial\psi^{2}}+D_{\perp}\frac{\partial^{2}}{\partial l\partial\psi}qP-\alpha D_{\perp}\frac{\partial}{\partial l}\left[\left(\partial_{\psi}+q\partial_{l}\right)q\right]P
-D_{\perp}\frac{\partial^{2}}{\partial l\partial v_{\parallel}}v_{\parallel}qq_{l}P
+\frac{1}{2}D_{\perp}\frac{\partial^{2}}{\partial v^{2}_{\parallel}}\left(v_{\parallel}q_{l}\right)^{2}P
-D_{\perp}\frac{\partial^{2}}{\partial\psi\partial v_{\parallel}}v_{\parallel}q_{l}P
+\alpha D_{\perp}\frac{\partial}{\partial v_{\parallel}}\left[v_{\parallel}\left(q\partial_{l}+\partial_{\psi}\right)q_{l}-v_{\parallel}q^{2}_{l}\right]P
\label{FPE}
\end{dmath}
\end{widetext}

\noindent Here, the parameter $\alpha\in\left[0,1\right]$ defines the stochastic integral and 
we introduced friction terms $\mathfrak{C}_{l}$, $\gamma$, $\mathfrak{C}_{v_{\parallel}}$, $\mathfrak{C}_{\theta}$, and $\mathfrak{C}_{\psi}$ that are required to preserve the total energy when the system is closed. 
$P$ is the distribution function (probability density) in $(l,v_{\parallel},\theta,\psi,\mu)$ space. 

\section{Definition of the Stochastic Integral and Time-Scale of Fluctuations}

\noindent The objective of this section is to show that the definition of the stochastic integral, i.e. the value of the parameter 

\begin{equation}
\alpha=\frac{1}{\tau}\int\displaylimits_{0}^{\tau}WdW=\frac{D_{\perp}^{-1}}{\tau}\int\displaylimits_{0}^{\tau}\left(\int\displaylimits_{0}^{t}\phi_{\theta}dt'\right)\phi_{\theta}dt\label{alpha}
\end{equation}

\noindent where $\tau$ is an arbitrary time and we used $dW_{\perp}=\Gamma_{\perp}dt=-D_{\perp}^{-1/2}\phi_{\theta}dt$, is ultimately determined by the ratio between time-scale of fluctuations $\tau_{f}$ and period of bounce motion $T$. Indeed, 
when a fluctuation starts to act, the electric field felt by a particle can be represented as:

\begin{equation}
\phi_{\theta}=\left(\frac{D_{\perp}}{\tau_{d}}\right)^{1/2}e^{-t/\tau_{d}}
\end{equation}

\noindent where $\tau_{d}$ is the typical decay time of the fluctuation. Integrating equation (\ref{alpha}) gives:

\begin{equation}
\alpha=\frac{1}{2}\left(1-e^{-\tau/\tau_{d}}\right)^{2}=\frac{1}{2}\left[1-exp\left(-\mathfrak{a}\frac{L\omega_{b}}{\delta\omega_{f}}\right)\right]^{2}
\end{equation}

\noindent Here, we used the fact that the decay time is given by:

\begin{equation}
\tau_{d}=\frac{\delta}{\sqrt{\frac{2}{m}T_{\perp}}}=\frac{\delta}{\mathfrak{a}\sqrt{\frac{2}{m}T_{\parallel}}}=\frac{\delta}{\mathfrak{a} v_{\parallel}}=\frac{2\pi}{\mathfrak{a} L\omega_{b}}
\end{equation}

\noindent with $\delta$ the typical size of fluctuation domains in the direction perpendicular to field lines, $\mathfrak{a}^{2}=T_{\perp}/T_{\parallel}$ the temperature anisotropy, $T_{\perp}$ and $T_{\parallel}$ the perpendicular and parallel temperatures respectively, $L$ the typical bounce orbit length, $\omega_{f}=2\pi/\tau_{f}$ the frequency of the fluctuations, and we set $\tau=\tau_{f}$. The limiting cases of interest are:
$\lim_{\omega_{b}/\omega_{f}\rightarrow 0}\alpha=0$ and $\lim_{\omega_{b}/\omega_{f}\rightarrow \infty}\alpha=1/2$.
Consequently, we expect fast fluctuations $\omega_{b}/\omega_{f}\rightarrow 0$ to be better represented by the choice $\alpha=0$. $\alpha=1/2$ will instead describe slow oscillations.

\section{Conservation of Bounce Action and Strength of Diffusion}

\noindent In order to relate emergence of temperature anisotropy and destruction of the second adiabatic
invariant, we need to understand the conditions under which the bounce action $J_{\parallel}$ is preserved for the system under examination. A direct calculation shows that:

\begin{dmath}
\frac{dJ_{\parallel}}{dt}=T\left( \phi_{\theta}\langle\left(\mu B+e\phi\right)_{\psi}\rangle-\langle\phi_{\theta}\rangle\left(\mu B+e\phi\right)_{\psi} -q\langle\phi_{\theta}\rangle\left(\mu B+e\phi\right)_{l}+\langle\phi_{\theta}\rangle mv_{\parallel}^{2}q_{l}\right)\label{dJdt3}
\end{dmath}

\noindent Here, $\langle\,\,\rangle/T$ is the bounce orbit average, with 
$T=\oint ds/\bar{v}_{\parallel}=2\pi/\omega_{b}$ the period of the bounce oscillation.

 Two important conclusions can be drawn from the result (\ref{dJdt3}). If the period of the bounce oscillation is small enough with respect to the time scale under consideration, $dJ_{\parallel}/dt$ can be neglected. The additional requirement of a small electric field $\phi_{\theta}$ results in even smaller variations of $J_{\parallel}$. However, when electromagnetic fluctuations are fast enough, i.e. their time scale $\tau_{f}$ is such that $\tau_{f}\gg T$, and when their amplitude is not negligible ($e\phi\gg H$), the second adiabatic invariant is destroyed. The latter condition correspond to a large diffusion parameter $D_{\perp}\sim \phi_{\theta}^{2}/\tau_{f}$ and the choice $\alpha=0$ in the FPE.

Taking the bounce orbit average, one recovers the classical result $\langle dJ_{\parallel}/dt \rangle=0$ of \cite{North}.
However, we remark that $dJ_{\parallel}/dt$ is well represented by its bounce orbit average only for slow and weak fluctuations $\phi_{\theta}$.

\section{Temperature Anisotropy}

\noindent In the following we investigate the numerical solution to equation (\ref{FPE}), focusing on 
the evolution of temperature profiles. We assume a Maxwell-Boltzmann distribution as initial condition and take Dirichlet boundaries. Axial symmetry $\partial_{\theta}=0$ is also used.
We will show that self-organization of a peaked profile goes together with preferential heating
of the normal temperature:

\begin{equation}
T_{\perp}\left(l,\theta,\psi\right)=\frac{\int{\mu BP dv_{\parallel}d\theta_{c}d\mu}}{\int{P dv_{\parallel}d\theta_{c}d\mu}}
\end{equation}

\noindent Indeed, when particles climb up the magnetic field as a result of inward diffusion, they preserve the magnetic moment $\mu$: the stronger the magnetic field, the larger the normal kinetic energy $\mu B$ stored in the cyclotron gyration. At the same time, particles with enough parallel kinetic energy $mv_{\parallel}^{2}/2$ have a high probability of exiting the boundaries (falling into the atmosphere of the planet) as the magnetic field is not strong enough to reverse their motion. With a stronger diffusion, the process speeds up and this translates in a systematic depopulation of particles with high parallel temperature:

\begin{equation}
T_{\parallel}\left(l,\theta,\psi\right)=\frac{\int{\frac{m}{2}v^{2}_{\parallel}Pdv_{\parallel}d\theta_{c}d\mu}}{\int{Pdv_{\parallel}d\theta_{c}d\mu}}
\end{equation} 

\noindent The scenario described above is reproduced numerically in figure \ref{fig1}, where two cases with different diffusion strength $D_{\perp}$ are compared (in the first case, $D_{\perp}$ is ten times the value it takes in the second one). 

\begin{figure}[h]
\hspace*{-0.4cm}\centering
\includegraphics[scale=0.35]{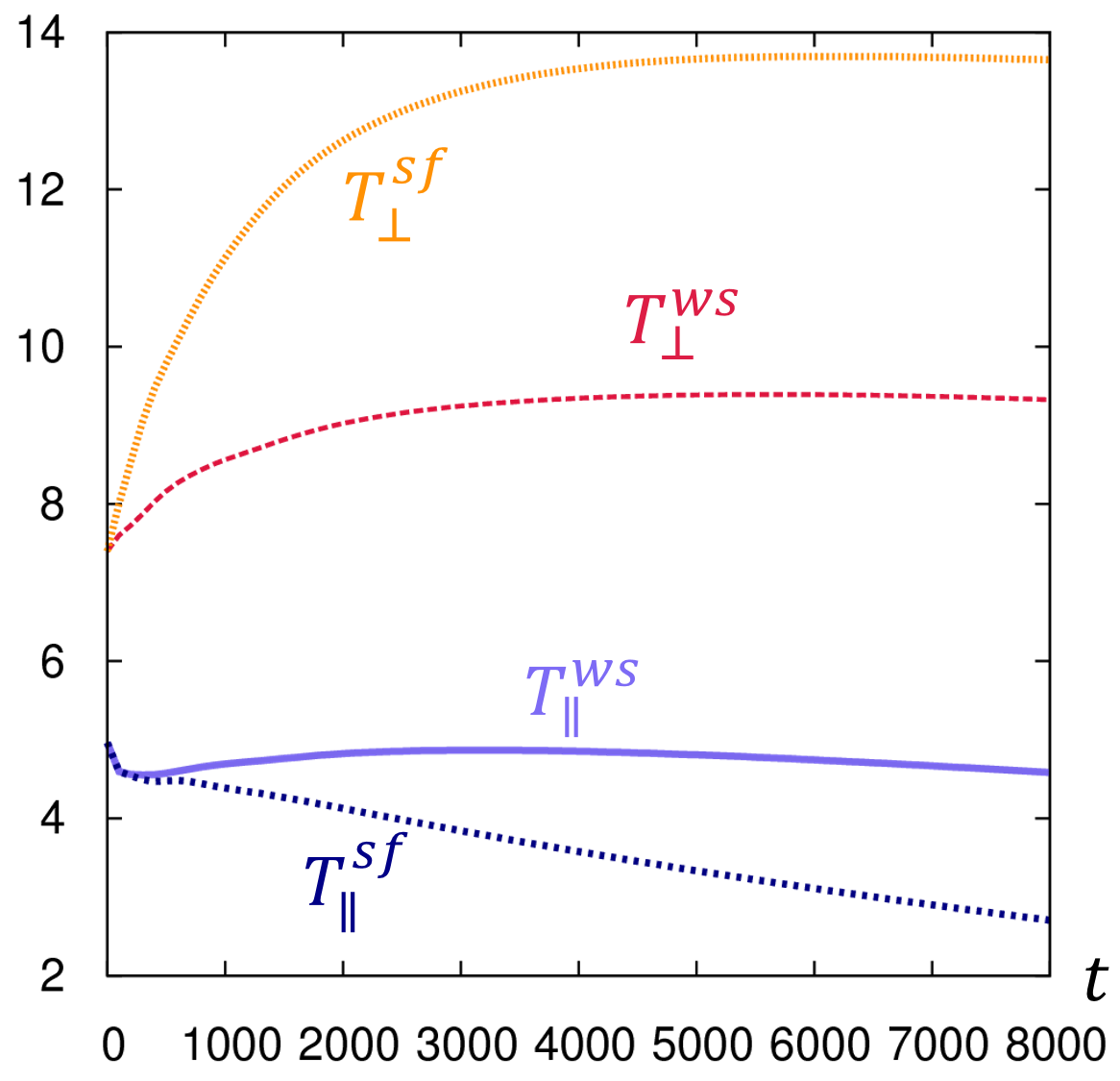}
\caption{\footnotesize Spatially averaged normal and parallel temperatures $T_{\perp}\left(eV\right)$ and $T_{\parallel}\left(eV\right)$ as a function of time $t(a.u.)$. $sf$ stands for strong and fast diffusion. $ws$ for weak and slow diffusion. The former case has a diffusion parameter $D_{\perp}$ ten times greater than the latter.}
\label{fig1}
\end{figure}

\noindent It is also instructive to look at the radial profiles of the temperatures $T_{\perp}$ and $T_{\parallel}$ in the two cases described above (figure \ref{fig2}). The preferential heating effect of the normal temperature due to inward diffusion is evident.  

\begin{figure}[h]
\hspace*{-0.0cm}\centering
\includegraphics[scale=0.35]{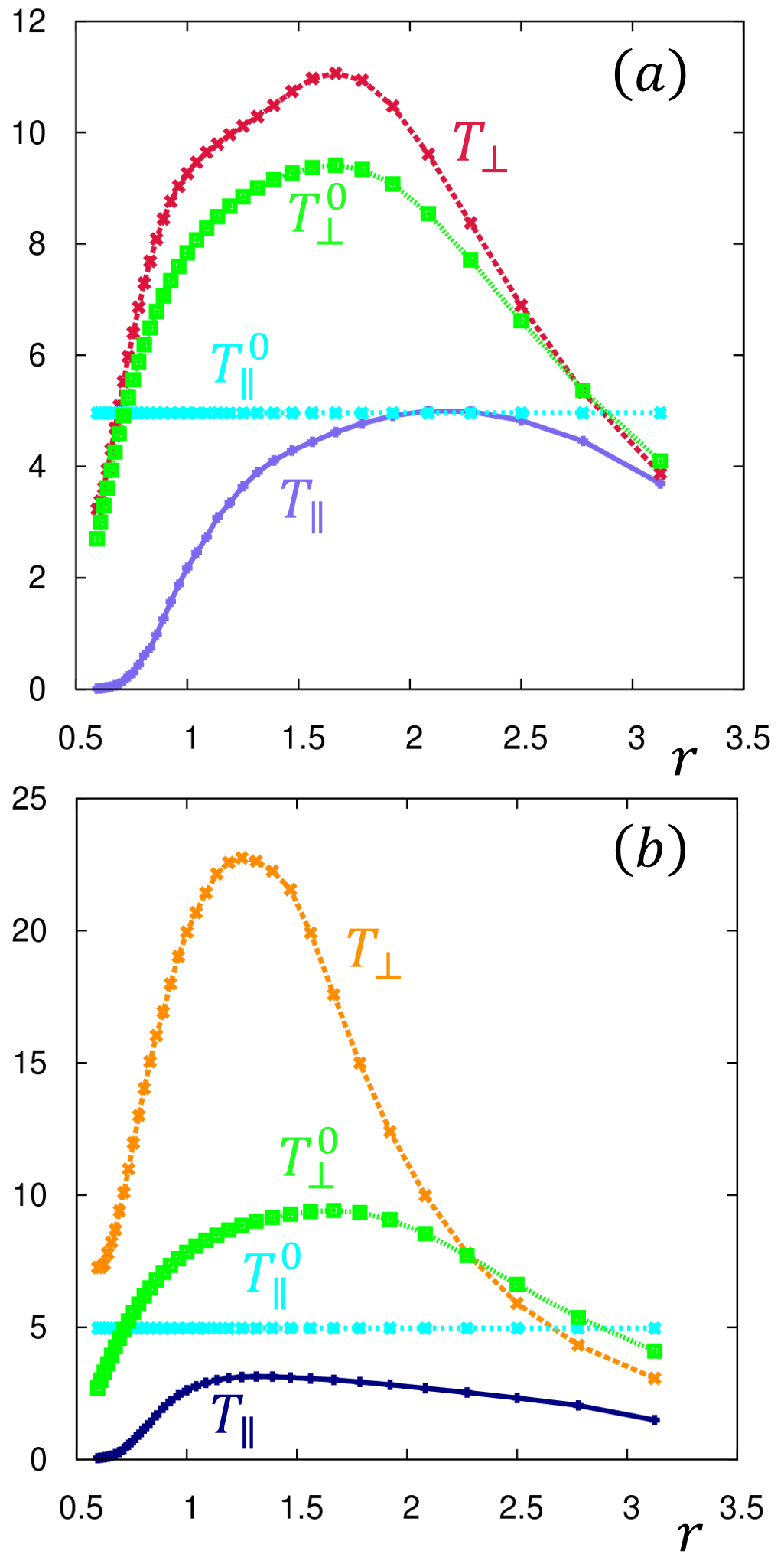}
\caption{\footnotesize (a): radial profiles of $T_{\perp}(eV)$ and $T_{\parallel}(eV)$ for the weaker diffusion parameter case. (b): radial profiles of $T_{\perp}(eV)$ and $T_{\parallel}(eV)$ for the stronger diffusion parameter case. The $0$ indicates the initial distribution. Here, $r$ is the radial coordinate of a cylindrical reference system $(r,z,\theta)$. $z$ is set to zero.}
\label{fig2}
\end{figure}

\section{Conclusion}

\noindent In this paper, continuing the efforts made to understand the physics of self-organization in magnetospheric plasmas \cite{YosVortex,YosRT1,YosFol,Sato2,Sato,Kaw}, we make important progress in the theoretical description of inward diffusion. Topological constraints dictate a natural reduction of phase space. However, there are several delicate problems, part of them yet unsolved, that have to be addressed when building a statistical mechanics on the foliation (existence of invariant measure, ergodicity, notion of entropy). Nevertheless, it is this reduced mechanics that is able to cast inward diffusion, and more generally self-organization phenomena, in the classical perspective of thermodynamics. Then, spatial gradients, temperature anisotropies, and preferential heating are just a consequence of a coordinate transformation connecting Cartesian coordinates to the foliation.   
 

\end{normalsize}

\end{document}